\documentclass[prl,superscriptaddress,twocolumn,longbibliography]{revtex4-1}

\usepackage{color}
\usepackage{mathtools}
\usepackage{tabularx}
\usepackage{graphicx}
\usepackage[hyperindex=true]{hyperref}

\definecolor{linkcol}{rgb}{0.2,0.2,0.6}

\hypersetup{bookmarksopen=true,
pdfmenubar=true, 
pdfhighlight=/O, 
colorlinks=true, 
pdfpagemode=None, 
pdffitwindow=true, 
linkcolor=linkcol, 
citecolor=linkcol, 
urlcolor=linkcol 
}

\newcolumntype{L}[1]{>{\raggedright\arraybackslash}p{#1}}
\newcolumntype{C}[1]{>{\centering\arraybackslash}p{#1}}
\newcolumntype{R}[1]{>{\raggedleft\arraybackslash}p{#1}}

\begin{document}

\title{Spin waves and three-dimensionality in the high-pressure antiferromagnetic phase of SrCu$_2$(BO$_3$)$_2$}

\author{Ellen Fogh}
\affiliation{Laboratory for Quantum Magnetism, Institute of Physics, 
Ecole Polytechnique F\'{e}d\'{e}rale de Lausanne (EPFL), CH-1015 
Lausanne, Switzerland}
\author{Gaétan Giriat}
\affiliation{Laboratory for Quantum Magnetism, Institute of Physics, 
Ecole Polytechnique F\'{e}d\'{e}rale de Lausanne (EPFL), CH-1015 
Lausanne, Switzerland}
\author{Mohamed E. Zayed}
\affiliation{Department of Physics, Carnegie Mellon University in Qatar, 
Education City, PO Box 24866, Doha, Qatar}
\author{Andrea Piovano}
\affiliation{Institut Laue-Langevin (ILL), 71 avenue des Martyrs, 38042 Grenoble, France}
\author{Martin Boehm}
\affiliation{Institut Laue-Langevin (ILL), 71 avenue des Martyrs, 38042 Grenoble, France}
\author{Paul Steffens}
\affiliation{Institut Laue-Langevin (ILL), 71 avenue des Martyrs, 38042 Grenoble, France}
\author{Irina Safiulina}
\affiliation{Institut Laue-Langevin (ILL), 71 avenue des Martyrs, 38042 Grenoble, France}
\author{Ursula B. Hansen}
\affiliation{Institut Laue-Langevin (ILL), 71 avenue des Martyrs, 38042 Grenoble, France}
\author{Stefan Klotz}
\affiliation{Sorbonne Université, UMR CNRS 7590, Institut de Minéralogie, de Physique des Matériaux et de Cosmochimie (IMPMC), Paris, France}
\author{Jian-Rui Soh}
\affiliation{Laboratory for Quantum Magnetism, Institute of Physics, 
Ecole Polytechnique F\'{e}d\'{e}rale de Lausanne (EPFL), CH-1015 
Lausanne, Switzerland}
\author{Ekaterina Pomjakushina}
\affiliation{Laboratory for Multiscale Materials Experiments, Paul Scherrer 
Institute, CH-5232 Villigen PSI, Switzerland}
\author{Fr\'{e}d\'{e}ric Mila}
\affiliation{Institute of Physics, 
Ecole Polytechnique F\'{e}d\'{e}rale de Lausanne (EPFL), CH-1015 Lausanne, Switzerland}
\author{Bruce Normand}
\affiliation{Laboratory for Quantum Magnetism, Institute of Physics, 
Ecole Polytechnique F\'{e}d\'{e}rale de Lausanne (EPFL), CH-1015 
Lausanne, Switzerland}
\affiliation{Laboratory for Theoretical and Computational Physics, 
Paul Scherrer Institute, CH-5232 Villigen-PSI, Switzerland}
\author{Henrik M. R\o nnow}
\affiliation{Laboratory for Quantum Magnetism, Institute of Physics, 
Ecole Polytechnique F\'{e}d\'{e}rale de Lausanne (EPFL), CH-1015 
Lausanne, Switzerland}

\begin{abstract}

Quantum magnetic materials can provide explicit realizations of paradigm models in quantum many-body physics. In this context, SrCu$_2$(BO$_3$)$_2$ is a faithful realization of the Shastry-Sutherland model (SSM) for ideally frustrated spin dimers, even displaying several of its quantum magnetic phases as a function of pressure. We perform inelastic neutron scattering (INS) measurements on SrCu$_2$(BO$_3$)$_2$ at 5.5 GPa and 4.5 K, observing spin waves that characterize the high-pressure antiferromagnetic phase. The experimental spectra are well described by linear spin-wave calculations on a SSM with an inter-layer interaction, which is determined accurately as $J_c = 0.053(3)$ meV. The presence of $J_c$ indicates the need to account for the three-dimensional nature of SrCu$_2$(BO$_3$)$_2$ in theoretical models, also at lower pressures. We find that the ratio between in-plane interactions, $J'/J = 1.8(2)$, undergoes a dramatic change compared to lower pressures that we deduce is driven by a sharp drop in the dimer coupling, $J$. Our results underline the wide horizons opened by high-pressure INS experiments on quantum magnetic materials.

\end{abstract}

\maketitle 

Frustrated magnetic interactions give rise to fascinating entangled quantum states, including valence-bond solids, quantum spin liquids, and spin ices \cite{savary2016}. Beyond the Heisenberg model on frustrated geometries such as the triangular and kagome lattices, exotic ground states, excitations, and quantum phase transitions (QPTs) are found in many systems where interactions compete in real space or in spin space. Despite multiple developments in both theoretical and numerical methods, the nature of many such states, their spectra, and the model phase diagrams remain only partially understood. Experimental studies of materials providing physical realizations are therefore an essential factor for our understanding of quantum magnetic systems.

\begin{figure}[t!]
	\centering
	\includegraphics[width = \columnwidth]{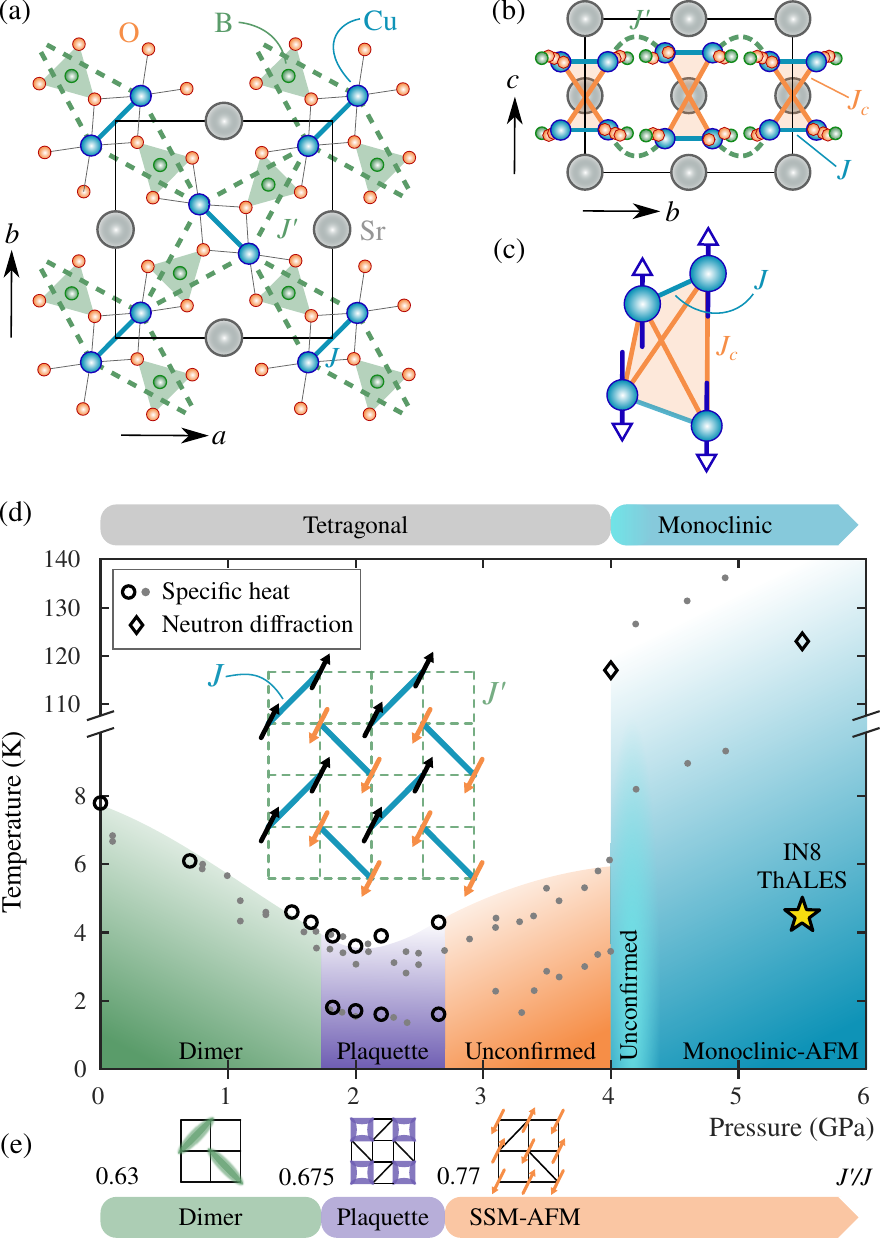}
	\caption{SrCu$_2$(BO$_3$)$_2$ and the Shastry-Sutherland model (SSM). 
	(a)-(b) Crystal structure of SrCu$_2$(BO$_3$)$_2$ shown in projection along $(0,0,1)$ (a) and $(1,0,0)$ (b). The dimer interaction, $J$ (solid blue line), inter-dimer interaction, $J'$ (dashed green line), and inter-layer interaction, $J_c$ (solid orange line), are indicated.
	(c) Illustration of the tetrahedron formed by two orthogonal dimers stacked along the $(0,0,1)$ direction.	
	(d) Phase diagram of SrCu$_2$(BO$_3$)$_2$ as a function of temperature and pressure, as deduced from specific-heat measurements (circles \cite{larrea2021} and dots \cite{guo2020}) and neutron diffraction (diamonds) \cite{haravifard2014,zayed2017}. The star marks the location in temperature and pressure of our INS experiments. The inset shows a schematic representation of the SSM.
	(e) Phase diagram of the SSM as a function of $J'/J$, as determined by multiple numerical methods \cite{koga2000,chung2001,lauchli2002,nakano2018,corboz2013,xi2023}.}
	\label{fig:Fig1}
\end{figure}

SrCu$_2$(BO$_3$)$_2$ is an ideally frustrated system composed of Cu$^{2+}$ dimers ($S = 1/2$ spin pairs) arranged orthogonally both in and between quasi-two-dimensional (2D) layers, as represented in Figs.~\ref{fig:Fig1}(a)-(c) \cite{kageyama1999b}. The in-plane network realizes the Shastry-Sutherland model (SSM) [inset Fig.~\ref{fig:Fig1}(d)], proposed \cite{shastry1981} for its exact dimer-singlet ground state at small $J'/J$ and governed by this ratio of competing interactions \cite{koga2000,chung2001,lauchli2002,corboz2013,nakano2018,xi2023}. SrCu$_2$(BO$_3$)$_2$ exhibits the dimer-singlet ground state at ambient pressure \cite{miyahara1999,kageyama1999b,kageyama2000a,kageyama2000b,gaulin2004,kakurai2005} and is well described by a SSM with $J'/J = 0.63$ and additional weak (3\%) Dzyaloshinskii-Moriya (DM) interactions \cite{cepas2001,miyahara2003,kodama2005,miyahara2005,shi2022}. Remarkably, the phase diagram of SrCu$_2$(BO$_3$)$_2$ under an applied hydrostatic pressure \cite{waki2007,zayed2017,sakurai2018}, shown in Fig.~\ref{fig:Fig1}(d), mirrors that of the SSM with increasing $J'/J$ [Fig.~\ref{fig:Fig1}(e)], and is studied extensively for its exotic QPTs \cite{guo2020,larrea2021,guo2023}. Above 4 GPa, where the tetragonal symmetry changes to monoclinic \cite{loa2005,haravifard2014,zayed2014b}, antiferromagnetic (AFM) order is found by neutron and X-ray diffraction \cite{haravifard2014,zayed2017,zayed2010} and phase-transition features are reported around 8 K \cite{guo2020} and above 120 K \cite{haravifard2014,zayed2014b,guo2020}. However, these high pressures pose a severe challenge to experiment and rather little is known about either the precise pressure regimes for the SSM-AFM and monoclinic-AFM phases or the possible magnetic interactions in either phase. 

Here we perform an inelastic neutron scattering (INS) experiment to characterize the monoclinic-AFM phase of SrCu$_2$(BO$_3$)$_2$. Working at 5.5 GPa and 4.5 K, we observe dispersive spin-wave excitations in 58 mg of sample. Linear spin-wave theory is appropriate to deduce the parameters of the minimal magnetic Hamiltonian. The in-plane interactions change dramatically from $J'/J = 0.63$ at ambient pressure to 1.8(2) at 5.5 GPa, which we ascribe to a strong reduction of $J$ arising from its near-critical bond angle. We also discover that it is necessary to include an inter-layer coupling, which we fit as $J_c = 0.053(3)$ meV, to explain the observed 2 meV splitting of otherwise degenerate branches in the low-energy part of the spectrum. We discuss the consequences of this 3D nature for the physics of SrCu$_2$(BO$_3$)$_2$. 

Our INS experiments used a 58 mg disc-shaped single-crystalline sample of SrCu$_2$(BO$_3$)$_2$, whose growth and characterization we summarize in Sec. S1A of the Supplemental Material (SM) \cite{supp}. The sample was oriented with ${\bf Q} = (q_h,q_k,0)$ in the horizontal scattering plane. Its small size is a challenge for INS, but recent developments in high-pressure techniques at the Institute Laue-Langevin (ILL) \cite{komatsu2014,klotz2019b,funnel2021} made this experiment possible. Using a Paris-Edinburgh press, the pressure was increased gradually to 5.5 GPa while the lattice parameters of SrCu$_2$(BO$_3$)$_2$ \cite{loa2005} were measured for pressure determination. The sample chamber was pre-cooled with liquid N$_2$ and cooling to 4.5 K was provided by a closed-cycle refrigerator. We performed two consecutive experiments at the IN8 and ThALES triple-axis spectrometers with the same loading. We used fixed momentum transfers $k_f = 2.66$ Å$^{-1}$ at IN8 and $k_f = 1.50$ Å$^{-1}$ at ThALES, providing respective instrumental resolutions (FWHM) of $1.05(1)$ and $0.167(4)$ meV. Scans with constant energy transfer, $E$, or constant momentum transfer, $\bf Q$, were performed with typical counting times of 5 min per point. In the SM \cite{supp} we provide detailed descriptions of the pressure loading procedure (Sec.~S1B), instrument configurations (Sec.~S1C), pressure determination (Sec.~S1D), and data treatment (Sec.~S1E). Supporting diffraction data were collected at IN3, also at the ILL, and this experiment is described in Sec.~S1F.

\begin{figure}
	\centering
	\includegraphics[width = \columnwidth]{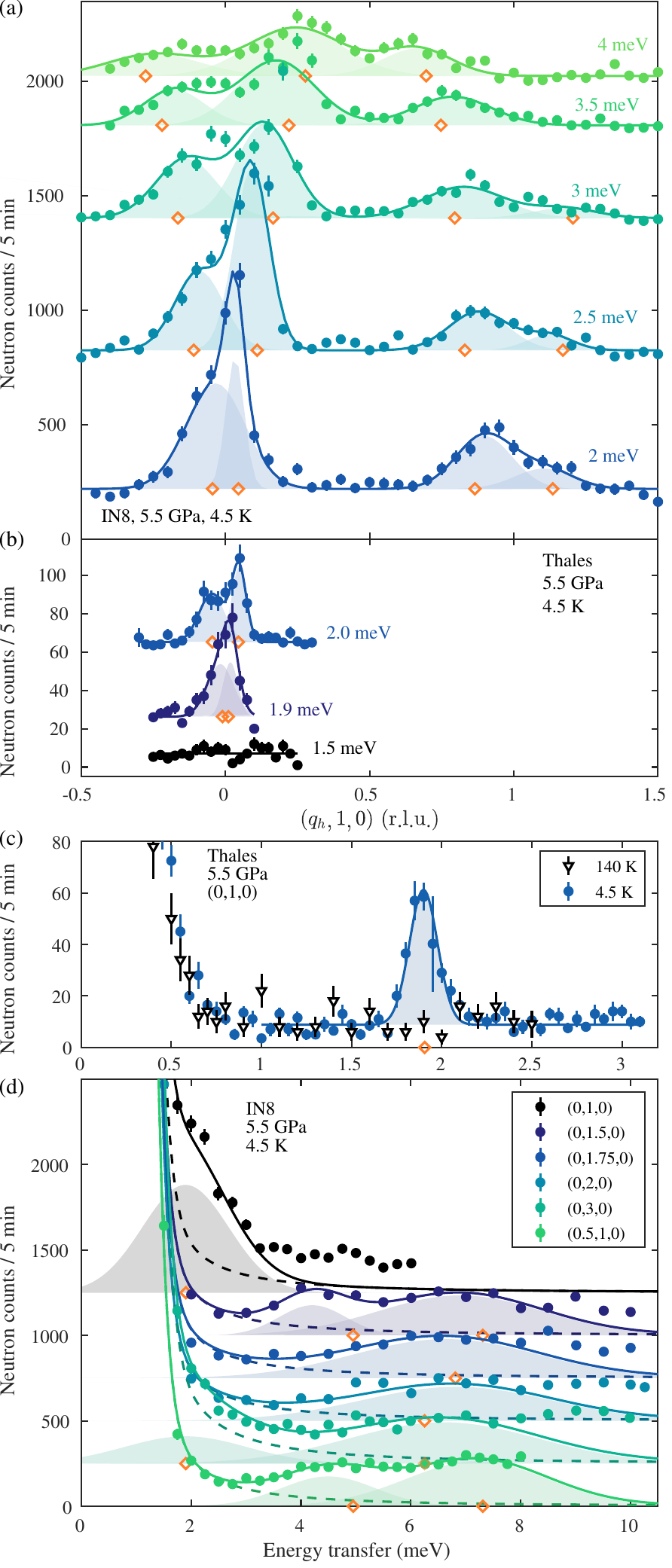}
	\caption{Selection of INS data collected at IN8 and ThALES at 5.5 GPa. (a)-(b) Constant-$E$ scans along ${\bf Q} = (q_h,1,0)$. (c)-(d) Constant-$\bf Q$ scans. Fits to determine the mode positions in $\bf Q$ and $E$ are illustrated with solid lines and the shaded areas represent the individual Gaussians. Dashed lines in panel (d) show the incoherent scattering contribution measured across the elastic line at $(0.7,1.3,0)$. Orange symbols mark the predicted mode centers obtained in a global parameter fit.}
	\label{fig:Fig2}
\end{figure}

\begin{figure*}
	\centering
	\includegraphics[width = \textwidth]{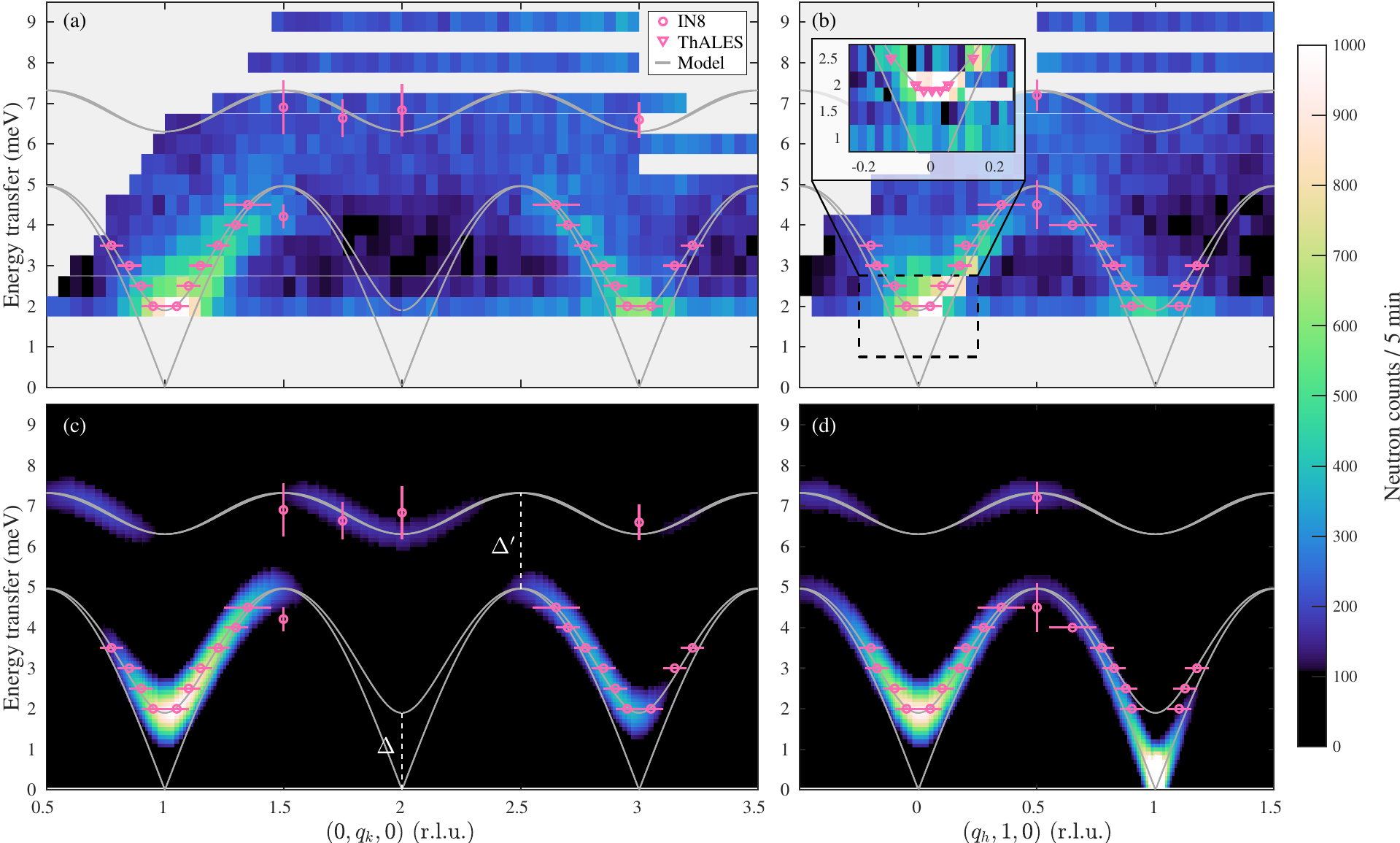}
	\caption{Measured and modeled excitation spectra of SrCu$_2$(BO$_3$)$_2$ in the monoclinic-AFM phase. 
	(a)-(b) INS intensity collected on IN8 at 5.5 GPa and 4.5 K, shown as a function of $E$ and $\bf Q$ along $(0,q_k,0)$ and $(q_h,1,0)$. The inset in panel (b) shows lower-energy data (dashed box) collected at ThALES around $(0,1,0)$.
	(c)-(d) Comparison with the spectrum calculated using an energy resolution of 1 meV. The splittings $\Delta$ and $\Delta'$ are indicated in panel (c). Mode positions determined from the INS data are shown by pink circles (IN8) and triangles (ThALES); the orientation of the error bar indicates whether a point originates from a constant-$E$ or constant-$\bf Q$ scan. The grey curves show the dispersion calculated from linear spin-wave theory.}
	\label{fig:Fig3}
\end{figure*}

A selection of constant-$E$ and constant-$\bf Q$ scans is shown in Fig.~\ref{fig:Fig2}, with data from both IN8 and ThALES. The width of the elastic line on IN8 allowed for access to energy transfers above approximately 2 meV, whereas at ThALES 0.5 meV was accessible. Because the in-plane lattice parameters of the monoclinic phase satisfy $a/b \simeq 0.999$ (with monoclinic angle $\beta \simeq 94^{\circ}$) \cite{haravifard2014}, we retain the tetragonal approximation and use $a^* = b^* = 2 \pi/a$. The color contours in Figs. \ref{fig:Fig3}(a)-(b) collect all neutron intensities measured respectively along $(0,q_k,0)$ and $(q_h,1,0)$ at constant $E$, and make clear that dispersive modes are present with minima at $(0,1,0)$ and $(1,1,0)$. At 1.5 meV and below, we no longer observe neutron intensity at $(0,1,0)$ [Figs.~\ref{fig:Fig2}(b) and \ref{fig:Fig2}(c)]. Above 5 meV it becomes difficult to discern sharp modes due to vanishing neutron intensity [Figs.~\ref{fig:Fig2}(a) and \ref{fig:Fig2}(d)]. In Figs.~\ref{fig:Fig2}(a)-(b) we fitted the data collected along $(q_h,1,0)$ with Gaussian line shapes centered respectively at $\pm \delta_h$ and $\pm \delta_h'$ from $(0,1,0)$ and $(1,1,0)$, allowing the peak widths and intensities to vary individually (Sec.~1E of the SM \cite{supp}). Similar fits to the data collected along $(0,q_k,0)$ yielded mode positions along both directions in $\bf Q$. The dispersion minimum at $(0,1,0)$ is resolved in a constant-$\bf Q$ scan from ThALES [Fig.~\ref{fig:Fig2}(c)]. A Gaussian fit yields an energy of 1.90(8) meV and the magnetic origin of the excitation is demonstrated by its absence at 140 K. In Fig.~\ref{fig:Fig2}(d), we show constant-$\bf Q$ scans from IN8 where subtraction of the incoherent contribution allowed the fitting of one or two Gaussian profiles at different $\bf Q$. The fitted mode positions are marked in Figs.~\ref{fig:Fig3}(a)-(b), where the relatively large error bars in the high-energy range reflect the difficulty in determining the excitation structure here. 

From the near-tetragonal nature of the high-pressure phase, we assume that the SSM continues to provide a minimal model for the observed spin-wave dispersion. The physics of the AFM phase is that the stronger $J'$ forces the spins to order ferromagnetically (FM) on the dimers and AFM between dimers, both in and out of plane [Fig.~\ref{fig:Fig1}(c)]. The 2D AFM phase has two doubly degenerate spin-wave branches, a Goldstone mode and an optical mode. Although we have no dispersion measurements in the $c$ direction, the observed in-plane scattering intensity [Figs.~\ref{fig:Fig3}(a)-(b)] motivates the introduction of an inter-layer interaction, $J_c$, between orthogonal dimers along the $(0,0,1)$ direction, as shown in Fig.~\ref{fig:Fig1}(c). This model has the following properties: (i) the degeneracy of the low-energy mode is lifted, with a splitting $\Delta = 8 S \sqrt{J' J_c}$ above the massless point; (ii) due to the AFM inter-layer stacking, the intensity of the Goldstone mode vanishes at ${\bf Q} = (0,1,0)$ and $(0,3,0)$, making the spectrum look gapped at these positions; (iii) there is an intensity asymmetry along $(0,q_k,0)$ around $q_k = 1$ and $3$, caused by the orthogonal arrangement of the dimers in the plane; (iv) the splitting between the low-energy and optical branches at the top of the former [e.g.~${\bf Q} = (0,1.5,0)$ and $(0.5,1,0)$] is $\Delta' \approx 2 S J$. 

We refine the model parameters based on the experimentally determined mode positions and intensities with assistance from \textsc{spinW} \cite{spinW}, a software implementation of linear spin-wave theory, to obtain $J' = 4.28(14)$ meV, $J = 2.36(26)$ meV (i.e.~$J'/J = 1.8(2)$), and $J_c = 0.053(3)$ meV. The predicted mode positions are shown in Fig.~\ref{fig:Fig2} as the orange symbols. The model spectra for an assumed instrumental energy resolution of 1 meV are shown in Figs.~\ref{fig:Fig3}(c)-(d), and provide good agreement with the measurements in Figs.~\ref{fig:Fig3}(a)-(b). In particular, the fitted spectrum confirms an AFM layer stacking, because FM stacking would result in Goldstone-mode intensity at $(0,1,0)$ and $(0,3,0)$, but not at $(1,1,0)$. We note that our deduced parameter values are not corrected for the quantum renormalization factor, $Z_c$, which to our knowledge has not been evaluated for the AFM phase of the SSM.

Currently the most refined estimates for the interaction parameters of SrCu$_2$(BO$_3$)$_2$ at ambient pressure are $J = 81.5$ K (7.0 meV) and $J' = 51.3$ K (4.4 meV, $J'/J = 0.63$) \cite{shi2022}. While there is no exact model for the increase of $J'/J$ with pressure, a linear decrease of both $J$ and $J'$, with $\Delta J/P$ falling 3-4 times faster than $\Delta J'/P$ \cite{zayed2017}, reproduces the $(H,P)$ phase diagram rather well in the dimer and plaquette phases \cite{shi2022}. \textit{Ab initio} calculations predict a discontinuity in the parameters on entering the SSM-AFM phase \cite{badrtdinov2020}, and at higher pressures no information for $J$ or $J'$ was available until now. Although it is unexpected that $J$ is suppressed so strongly, it is not implausible. Its size and sign are highly sensitive to the Cu-O-Cu bond angle following the Goodenough-Kanamori rules \cite{goodenough1955,kanamori1959}, especially when this angle is close to 95$^{\circ}$ \cite{mizuno1998}. Measurements in SrCu$_2$(BO$_3$)$_2$ [Figs.~\ref{fig:Fig1}(a)-(b)] indicate a decrease from $98^{\circ}$ at ambient pressure to $94^{\circ}$ at 3.7 GPa \cite{zayed2010}. The $J'$ superexchange path is through a BO$_3$ unit, which remains robust with pressure and locked in position by a mirror plane, becoming free to rotate out of the plane only beyond the monoclinic transition. Our results therefore quantify all of these considerations. 

Turning to $J_c$, we note first that the system is much softer along the ${\hat c}$ direction than in the plane, with the $c$ lattice parameter reduced by around 6\% at 5.5 GPa, whereas $a$ and $b$ are reduced by less than 1\% \cite{loa2005,haravifard2014}. The buckling of the Cu-BO$_3$ planes shown in Fig.~\ref{fig:Fig1}(b) leads to two independent $J_c$ parameters \cite{miyahara2003}. However, the in-plane INS spectrum is sensitive only to their combined effect, and thus we fit a single effective $J_c$; further details are presented in Sec.~S2A of the SM \cite{supp}. In this quasi-2D system, the splitting $\Delta$ is a steep function of $J_c$ and in our INS experiment $\Delta = 1.90(8)$ meV is measured to high accuracy, making our estimate of $J_c = 0.053(3)$ meV at 5.5 GPa very precise. Although this small value reflects the absence of significant inter-layer superexchange paths, it is still thought that $J_c$ has stronger contributions from superexchange than from direct exchange, making its evolution under pressure difficult to predict. Nevertheless, the ambient-pressure $J_c$ is unlikely to be much larger than 0.05 meV. An indirect estimate of 0.7 meV, an order of magnitude above our result, was obtained as a 10\% correction to the high-temperature magnetic susceptibility \cite{miyahara2000,miyahara2003}, but was presumably a consequence of strong finite-size effects inherent to early numerical studies (since remedied by more modern technology \cite{wietek2019}).

Although a Heisenberg SSM with an inter-layer coupling captures all the measured features of the spin-wave dispersion in the high-pressure phase, we draw attention to two key points. First, SrCu$_2$(BO$_3$)$_2$ has weak DM interactions at ambient pressure \cite{miyahara2003,kodama2005,shi2022}. These terms also act to gap the Goldstone mode, but the DM interaction required to create a gap of 1.9 meV in a 2D SSM is $0.7 J'$, as we show in Sec.~2B of the SM \cite{supp}, making this a very unlikely mechanism to explain the INS spectrum. In our minimal model, we absolutely do not claim that DM interactions are no longer present, only that we cannot resolve them. In Sec.~2B we show also that even 10\% DM terms in the lower-symmetry monoclinic phase of SrCu$_2$(BO$_3$)$_2$ would provide only a minor correction to the Heisenberg interaction parameters. Second, we stress that linear spin-wave theory does not account for two-magnon scattering, possible two-magnon bound states, or any other effects of quantum fluctuations caused by the frustration remaining in the system, all of which, as we summarize in Sec.~2C of the SM \cite{supp}, could contribute to the broad and weak neutron intensities observed above 5 meV in Figs.~\ref{fig:Fig2}(d) and \ref{fig:Fig3}(a)-(b).

The confirmation of a small $J_c$ is essential information for assessing the suitability of a 2D SSM to describe SrCu$_2$(BO$_3$)$_2$. The fact that the dimer ($J$) bonds are forced into a FM configuration in the AFM phase makes the four $J_c$ bonds in an inter-layer tetrahedron fully unfrustrated [Fig.~\ref{fig:Fig1}(c)], and hence they contribute an energy of order 1 K per site to the stability of a 3D AFM phase. In the dimer-singlet phase, by contrast, these bonds are simply ineffective. Concerning the plaquette phase, it was shown rather early that its very existence depends on having only a weak $J_c$ \cite{koga2000b}, and the propensity of $J_c$ to destabilize the plaquette phase has been used recently to extract a value $J_c \approx 0.2$ meV \cite{vlaar2023}. Our definitive $J_c$ can be used to refine the estimates of intra-layer energetics and their pressure evolution on which that estimate was based. Another delicate issue is whether the plaquette singlets form around the squares that contain the dimer bond (``full plaquette'') \cite{zayed2017} or the squares that do not (``empty plaquette'') \cite{corboz2013}. The minimal $J_c$ we consider should not affect plaquette formation \cite{vlaar2023}, but too little is known about the 3D plaquette order to make a statement on whether more complex inter-layer coupling could do so. Finally, because the plaquette-AFM transition of the 2D SSM has been proposed as a candidate deconfined quantum critical point (DQCP), and SrCu$_2$(BO$_3$)$_2$ as a candidate material for its observation \cite{lee2019,yang2022,cui2023}, the effects of inter-layer coupling should definitely be included as part of evaluating such a scenario. 

Finally, we remark that our results do not provide an immediate understanding of the reported phase transitions at 8 K \cite{guo2020} and 120 K \cite{haravifard2014,zayed2014b,guo2020}, the parameters of our quasi-2D model being more likely to favor a single ordering transition at an intermediate temperature (of order $J'/2$ \cite{yasuda2005}). We performed detailed diffraction studies of the $(0,1,0)$ magnetic Bragg peak, as described in Sec.~S1F of the SM \cite{supp}, finding an unconventional thermal evolution with a characteristic temperature of 88 K. While one may speculate about the onset of 2D and of true 3D order, these contradictions underline the need for more systematic combined studies of the structure and magnetism at and above 3 GPa.  

In conclusion, we have performed INS on a single crystal of 58 mg to measure spin waves in the monoclinic-AFM phase of SrCu$_2$(BO$_3$)$_2$ at a pressure of 5.5 GPa. The observed spectra are well described by a Shastry-Sutherland model with an inter-layer coupling. Consequently, when modelling the physics of the SrCu$_2$(BO$_3$)$_2$ phase diagram, it may be necessary to consider the three-dimensionality. Our work highlights the importance of the technical capability to perform INS under extreme conditions of high pressure and low temperature for the discovery and investigation of quantum magnetic states.

{\it{Acknowledgements}} -- We are very grateful to C. Payre and J. Maurice for their technical assistance with the high-pressure equipment at the ILL. We acknowledge the financial support of the European Research Council through the Synergy network HERO (Grant No.~810451) and of the Swiss National Science Foundation through Project Grant No.~188648. We thank the ILL for the allocation of neutron beam time for this study on the IN8, ThALES, and IN3 spectrometers. Collected data are available as Refs.~\cite{IN8_Thales_doi,IN3_doi}. This publication was made possible in part by the generous support of the Qatar Foundation through the Seed Research Funding Program of Carnegie Mellon University in Qatar. The statements made herein are solely the responsibility of the authors.


%

\end{document}